\documentclass[%
reprint,
superscriptaddress,
nofootinbib,
 amsmath,amssymb,
 aps,
]{revtex4-1}

\usepackage{graphicx}
\usepackage{dcolumn}
\usepackage{bm}
\usepackage{color}

\begin{document}

\preprint{APS/123-QED}

\title{Extracting the cutoff frequency in the gravitational-wave spectrum of black hole-neutron star mergers}

\author{Kyohei Kawaguchi}
\affiliation{Max Planck Institute for Gravitational Physics (Albert Einstein Institute), Am M\"{u}hlenberg 1, Potsdam-Golm, 14476, Germany}
\affiliation{Center for Gravitational Physics, Yukawa Institute for Theoretical Physics, Kyoto University, Kyoto 606-8502, Japan}
\author{Koutarou Kyutoku}
\affiliation{Theory Center, Institute of Particle and Nuclear Studies, KEK, Tsukuba 305-0801, Japan}
\affiliation{Department of Particle and Nuclear Physics, the Graduate University for Advanced Studies (Sokendai), Tsukuba 305-0801, Japan}
\affiliation{Interdisciplinary Theoretical Science (iTHES) Research Group, RIKEN, Wako, Saitama 351-0198, Japan}
\affiliation{Center for Gravitational Physics, Yukawa Institute for Theoretical Physics, Kyoto University, Kyoto 606-8502, Japan}
\author{Hiroyuki Nakano}
\affiliation{Faculty of Law, Ryukoku University, 67 Fukakusa Tsukamoto-cho, Fushimi-ku, Kyoto 612-8577, Japan}
\affiliation{Center for Gravitational Physics, Yukawa Institute for Theoretical Physics, Kyoto University, Kyoto 606-8502, Japan}
\author{Masaru Shibata}
\affiliation{Center for Gravitational Physics, Yukawa Institute for Theoretical Physics, Kyoto University, Kyoto 606-8502, Japan}
\date{\today}

\newcommand\apjl{ApJL}%
\newcommand{\rednote}[1]{{\color{red} (#1)}}

\begin{abstract}
	The location of the cutoff in the gravitational-wave spectrum of black hole-neutron star mergers is correlated strongly with the neutron-star radius for the case that the neutron star is disrupted by the black hole during the merger. However, the modulation which appears in the spectrum due to the mode mixing makes it difficult to measure the cutoff frequency if gravitational waves are observed from inclined direction or the binary is precessing. In this letter, we show that the cutoff frequency can be measured even in such situations with a method we have recently proposed to reconstruct the face-on waveforms only from the strain observed from a particular direction. We show that the systematic error in the measurement of the neutron-star radius can be reduced to $\lesssim 5\%$ for the case that tidal disruption of the neutron star occurs significantly.
\begin{description}
\item[PACS numbers]04.30.−w, 04.40.Dg, 04.25.D- 
\end{description}
\end{abstract}

\pacs{Valid PACS appear here}
\maketitle


\section{Introduction}
 Gravitational waves from black hole-neutron star mergers are among the most promising target for ground-based-detectors~\citep{2017PhRvL.118v1101A,2015CQGra..32b4001A,2012CQGra..29l4007S}, and they are expected to be detected in the not-so-far future~\citep{2010CQGra..27q3001A,2015ApJ...806..263D,2015MNRAS.448..928K}. The waveforms will contain rich information of the compact objects and physics in the extreme environment, e.g., strong gravitational field and high density and so on. In particular, gravitational waves contain the information of the neutron-star mass and radius that has a great impact on the study of nuclear physics via constraining the uncomprehended equation of state of the neutron-star matter. The black hole-neutron star mergers are also expected to be sources of various electromagnetic transients~\citep{2016ARNPS..66...23F}. Observations of the electromagnetic signal associated with the merger are useful to determine the host galaxy of the events. This enables us to measure the sky location and the redshift of the merger event accurately.
 
	To extract physical information from the waveforms and to understand the associated electromagnetic transients, quantitative prediction of the merger dynamics is necessary. The numerical-relativity simulation is a unique method to predict the late inspiral and merger stages quantitatively. Many numerical-relativity simulations have been performed for black hole-neutron star mergers in the last decade, and we have learned their dynamics quantitatively~\cite{2006PhRvD..74l1503S,2007CQGra..24S.125S,2008PhRvD..77h4015S,2008PhRvD..77h4002E,2008PhRvD..78j4015D,
   2012PhRvD..85l7502S,2009PhRvD..79d4024E,2010CQGra..27k4106D,2010PhRvD..82d4049K,
  2011PhRvD..84f4018K,2013CQGra..30m5004L,2015PhRvD..92d4028K,2010PhRvL.105k1101C,
   2012PhRvD..85f4029E,2012PhRvD..86h4026E,
  2013ApJ...776...47D,2014PhRvD..90b4026F,
   2011PhRvD..83b4005F,2013PhRvD..87h4006F,
   2015PhRvD..92f4034K,2015PhRvD..92b4014K,2015ApJ...806L..14P}. The correlation between the neutron-star radius and the {\it cutoff frequency} in the gravitational-wave spectrum is one of the most spectacular findings~\citep{2009PhRvD..79d4030S,2010PhRvD..82d4049K,2011PhRvD..84f4018K,2015PhRvD..92h1504P}. If the black-hole tidal force exceed the self-gravitational force of a neutron star during the inspiral stage, the neutron star can be disrupted by the tidal force of the black hole. In such a case, the amplitude of gravitational waves is strongly suppressed after the neutron star is tidally disrupted, and this suppression appears as damping in the higher frequency part of the gravitational-wave spectrum. The location of the damping in the spectrum, i.e.,  the cutoff frequency, is correlated strongly with the neutron-star radius. Hence, the measurement of the cutoff frequency along with the other binary parameters, such as the mass and the spin of each object, will constrain the radius of the neutron star. On the other hand, the precise measurement of the gravitational-wave phase evolution enables us to determine the parameter related to the neutron-star radius, such as the tidal deformability of the neutron star~\cite{2008PhRvD..77b1502F,2011PhRvD..83h4051V,2010PhRvD..81h4016D,2012PhRvD..85l4034B,2014PhRvD..90l4037B,2015PhRvL.114p1103B,2012PhRvD..85d4061L,2014PhRvD..89d3009L}.
The measurement of the cutoff frequency as well as the phase evolution enables us to perform the consistency check between the neutron-star parameters for a single gravitational-wave event.

	For most of the previous studies, the cutoff frequency was studied only considering the waveforms observed along the orbital axis. However, the binary would be generally observed from an inclined direction from the orbital axis. Moreover, the inclination angle can be a function of time if the orbital plane precesses due to the misalignment of the black-hole spin and the orbital angular momentum via the spin-orbit interaction~\citep{1994PhRvD..49.6274A,1995PhRvD..52..821K}. In such situations, not only the dominant wave components, of which frequency is twice the frequency of the orbital motion, but also other components with different frequencies significantly contribute to the observed waveforms. This mixing of the modes with different frequencies causes the modulation in the amplitude of the gravitational-wave spectrum, which makes the location of the cutoff obscure~\citep{2015PhRvD..92b4014K}. Therefore, the direct measurement of the cutoff frequency for the observed waveforms is not a trivial issue for generic situations. 
	
	In this letter, we propose a method to measure the location of the cutoff  in the gravitational-wave spectrum for the case that the binary is observed from an inclined direction. We show the systematic error in the measurement of the neutron-star radius by our method when the gravitational-wave detector noise is ignored.  Throughout this letter, we refer to the waveforms observed along the orbital axis as the face-on waveforms. For the precessing cases, the face-on waveforms are defined by the waveforms observed along the instantaneous orbital axis employing the definition of Ref.~\citep{2011PhRvD..84b4046S,2012PhRvD..86j4063S}. 

\section{Method}
	The analysis in this letter is based on the method introduced in our recent paper~\citep{2017arXiv170507459K}. This method allows us to approximately reconstruct the face-on waveforms only from the waveform strain observed from a particular direction. The brief outline of the method is as follows: First, the orbital phase is approximately extracted from the phase of the observed waveforms by fitting a smooth function. Second, the Fourier spectrum with respect to the orbital phase, which we call the mode spectrum, is computed from the strain. Third, the contribution from the dominant modes is extracted from the mode spectrum and it is re-transformed into the time domain data. Finally, the face-on waveforms are reconstructed from those extracted dominant-mode waveforms. The contribution of the sub-dominant modes is removed from the strain by this procedure, and hence, the modulation in the gravitational-wave spectrum is removed by employing the reconstructed waveforms. Our purpose of this letter is to demonstrate that our method enables us to measure the cutoff frequency and to constrain the neutron-star radius within an acceptably small systematic error. For this purpose, we do not consider the noise in the strain in this letter.
	
	We have improved the fitting function from the previous work~\citep{2017arXiv170507459K} in two ways. We eliminate the logarithmic terms in the fitting function in the inspiral stage because they do not improve the accuracy of the fit. The fitting function for the post-merger stage is slightly changed and the junction condition between the inspiral stage and the post-merger stage is strengthened to guarantee the ${\cal C}^2$ condition at $t=t_0$ of Eq.~(11) in Ref.~\citep{2017arXiv170507459K}. We newly set $t_0$ as the fitting parameter, which was fixed in the previous paper. We also change the extraction procedure to apply the window function of Eq.~(16) in Ref.~\citep{2017arXiv170507459K} individually for each $|m|=2$ mode while both $m=\pm2$ were extracted at once in the previous work.
	
	We use the waveforms obtained in the numerical-relativity simulations performed in Refs.~\citep{ 2015PhRvD..92d4028K,2015PhRvD..92b4014K}. In these simulations, the neutron-star mass is fixed to be $1.35\,M_\odot$. We employ the models with the mass ratio of the black hole to the neutron star and the dimensionless spin parameter of the black hole are $5$ and $0.75$, respectively. By contrast, the neutron-star radius and the initial black-hole spin orientations, $i_{\rm tilt}$, are varied systematically. Here, $i_{\rm tilt}$ is defined as the angle between the black-hole spin and the orbital angular momentum at the initial time of the simulation. We generate waveform strains observed from various directions employing the waveform data up to $l=4$ components of spin-weighted spherical harmonic expansion. We choose $\theta=(0,$ $\pi/12,$ $\pi/6,$ $\pi/3,$ $5\pi/12$$)$ and $\phi=(0,$ $\pi/2,$ $\pi,$ $3\pi/2$$)$ for each $\theta$ as fiducial observer directions. Here, two angles, $\theta$ and $\phi$, denote the direction of the fiducial observer from the binary in Cartesian coordinates in which the $x$ and $z$ axes are taken along the separation vector from the black hole to the neutron star and the total angular momentum at the initial time of simulation, respectively. We note that we only consider the case that $\theta\le5\pi/12$ because the denominator of Eq.~(10) in~\citep{2017arXiv170507459K} can be zero for the case that $\theta\approx\pi/2$ and our method cannot be used for such a case. Since binaries with smaller values of $\theta$ are more likely to be detected in the gravitational-wave observation, this restriction may not be severe \citep{2011CQGra..28l5023S}.

	The cutoff frequency in the spectrum is defined in the same way as in Ref.~\citep{2015PhRvD..92h1504P}: First, we search the frequency $f_0$ at which $f^2|{\tilde h}\left(f\right)|$ becomes maximum, where ${\tilde h}(f)$ is the Fourier spectrum of the strain. Then, we define the cutoff frequency of the spectra, $f_{\rm cut}$, as the frequency that the amplitude of the spectra becomes ${\tilde h}\left(f_0\right)/e$ for the first time at $f_0<f_{\rm cut}$. We confirm that the cutoff frequency thus defined agrees with the quasinormal-mode frequency of the remnant black hole within $\sim10\%$ for binary black holes.

\section{Results}
\begin{figure}
	\begin{center}
		\includegraphics[width=90mm]{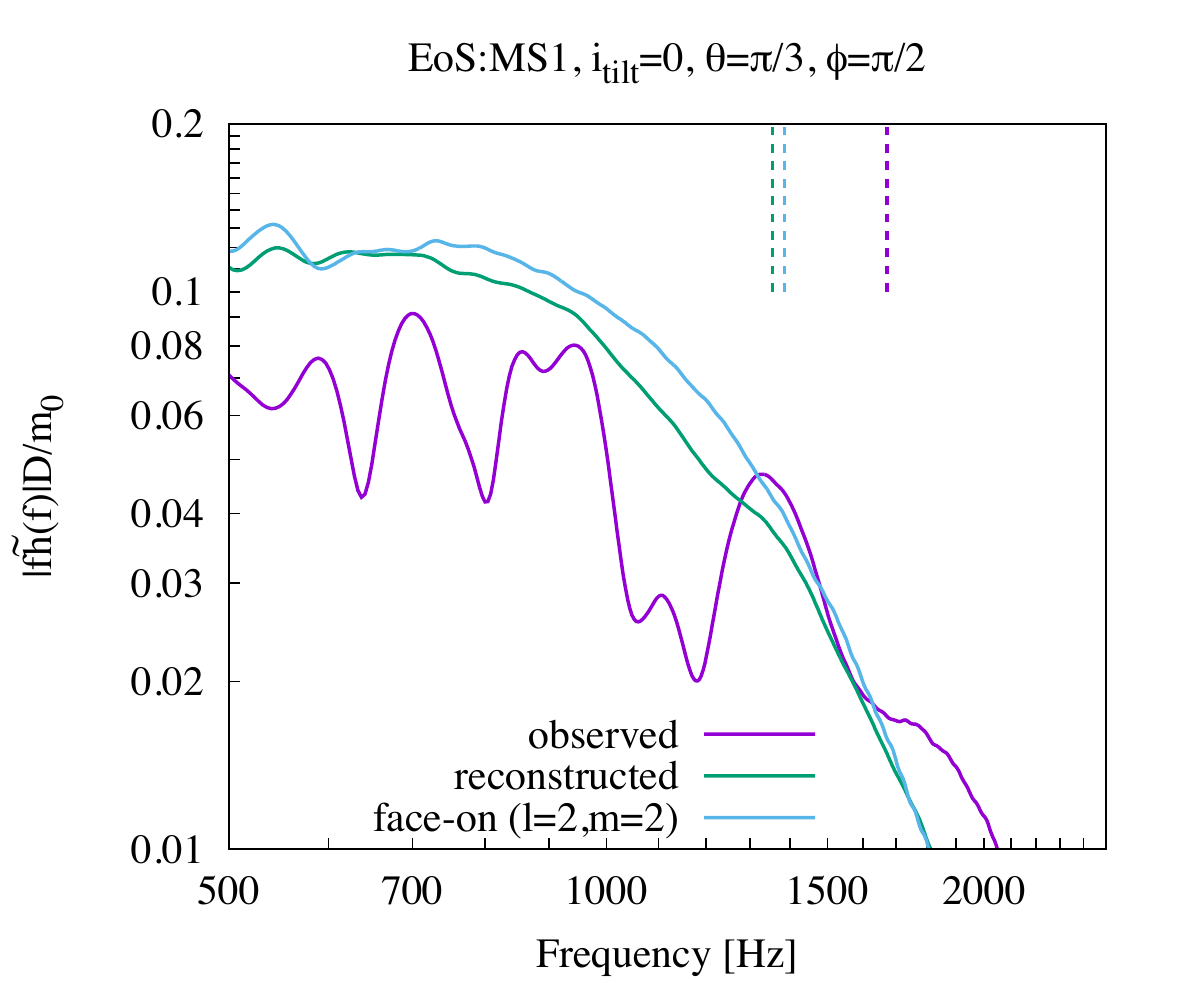}
	\end{center}
	\caption{The comparison between the gravitational-wave spectra calculated from the strain observed from a particular observer direction, the waveforms reconstructed from the strain by our method, and the face-on waveforms. Here, $m_0$ and $D$ are the total mass of the binary and the hypothetical distance to the binary in geometrical units, respectively. The vertical dashed lines denote the cutoff frequency of each spectrum.}
	\label{fig:gws}
\end{figure}

	The extraction of the face-on waveforms by the mode decomposition method is crucial for measuring the cutoff frequency from the observed strain. In Fig.~\ref{fig:gws}, we compare the spectra calculated from the waveforms observed from a particular observer direction, the waveforms reconstructed by our method, and the face-on waveforms directly calculated from numerical-relativity data. The spectra of observed strains are modulated, and the modulation produces fake cutoffs in them for both non-precessing and precessing cases. On the other hand, such modulation disappears in the spectra of the reconstructed waveforms, and the spectrum agrees nicely with the one of the face-on waveforms. Indeed, we find that the cutoff frequency measured from the reconstructed waveforms agrees with that of the face-on waveforms within $\sim 2\%$ for this example, while the one measured from the observed strain deviates from them. This result clearly shows that the reconstruction of the face-on waveforms is crucial to measure the cutoff frequency.

\begin{figure*}
	\begin{center}
		\includegraphics[width=80mm]{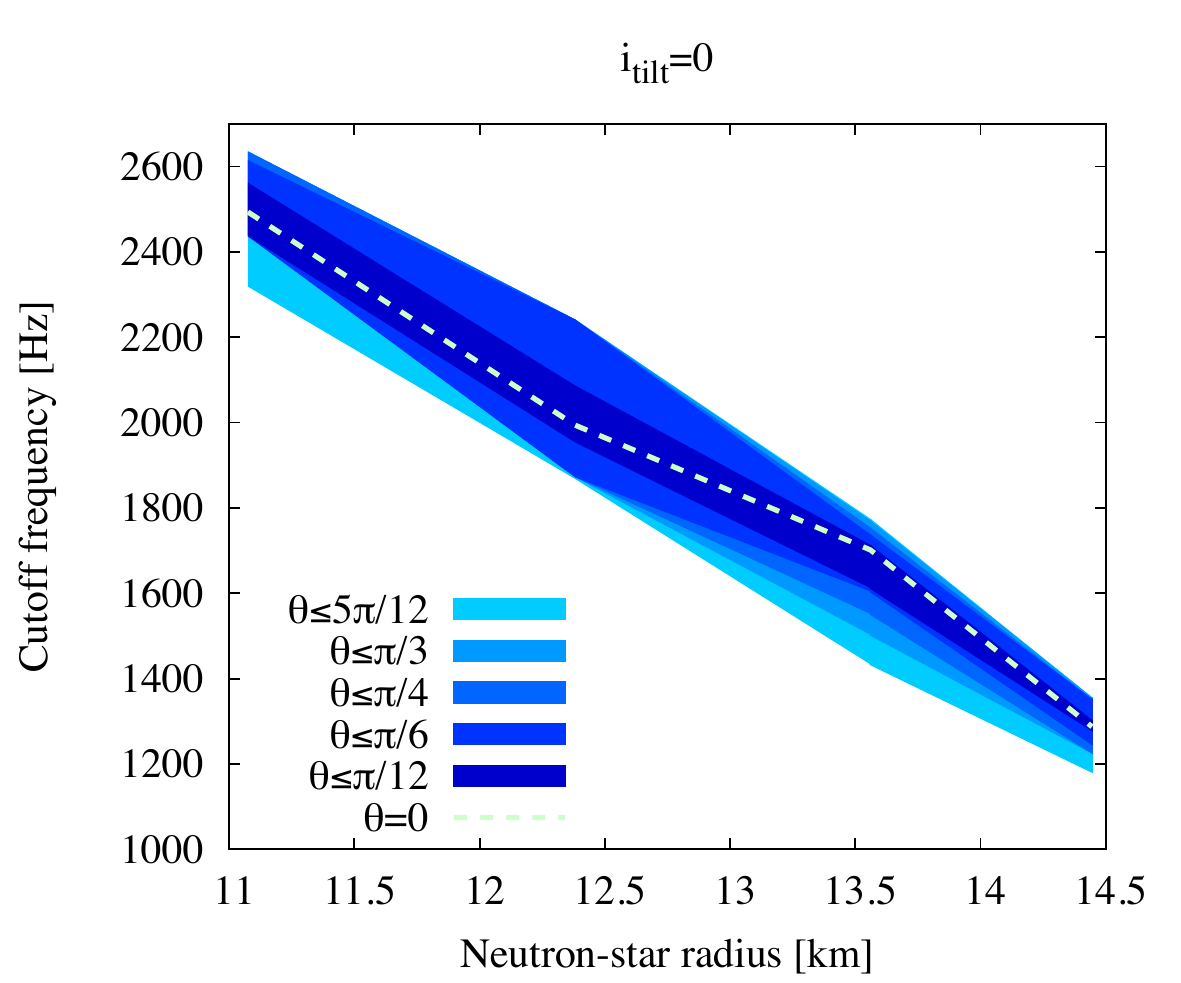}
		\includegraphics[width=80mm]{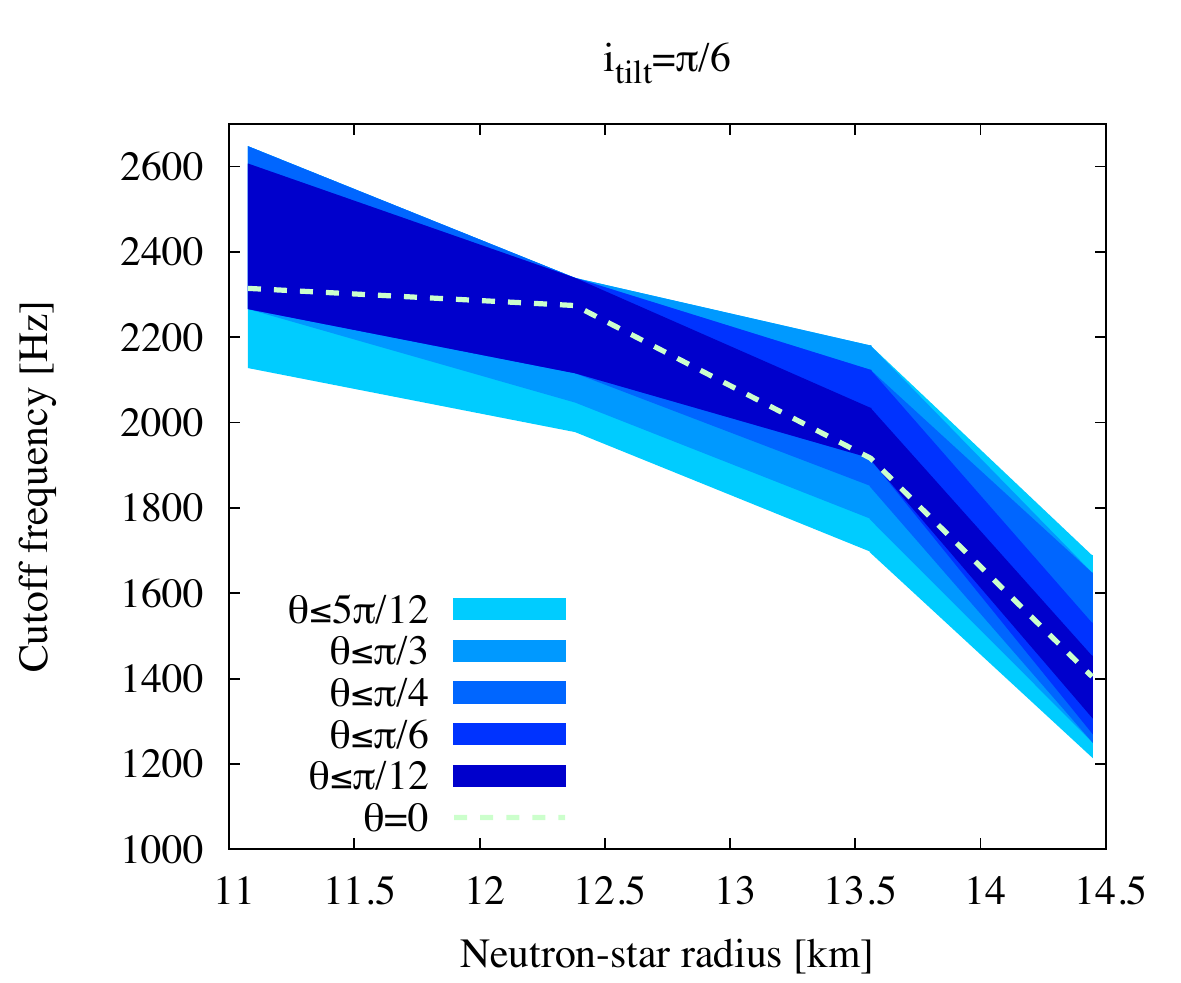}\\
		\includegraphics[width=80mm]{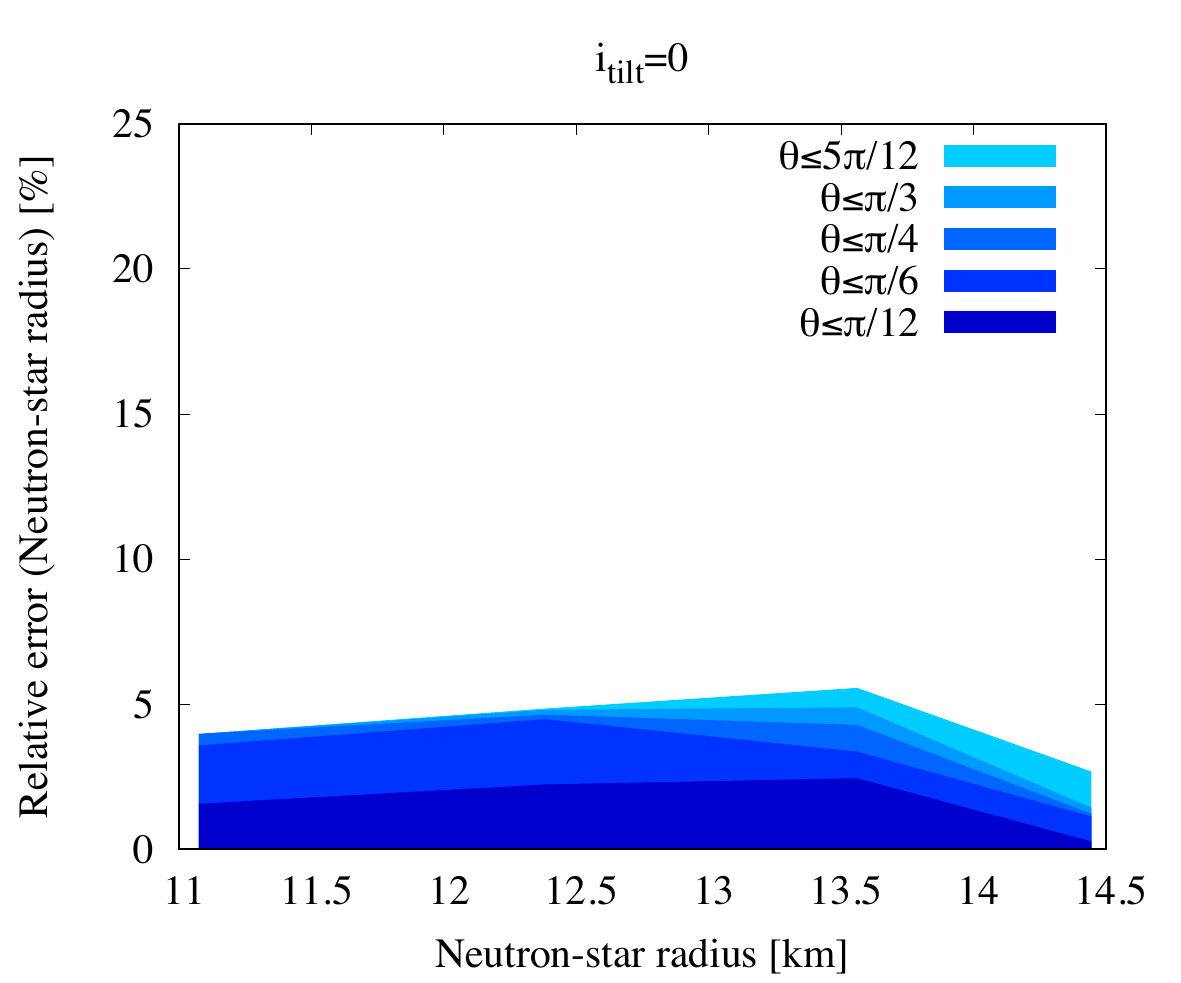}
		\includegraphics[width=80mm]{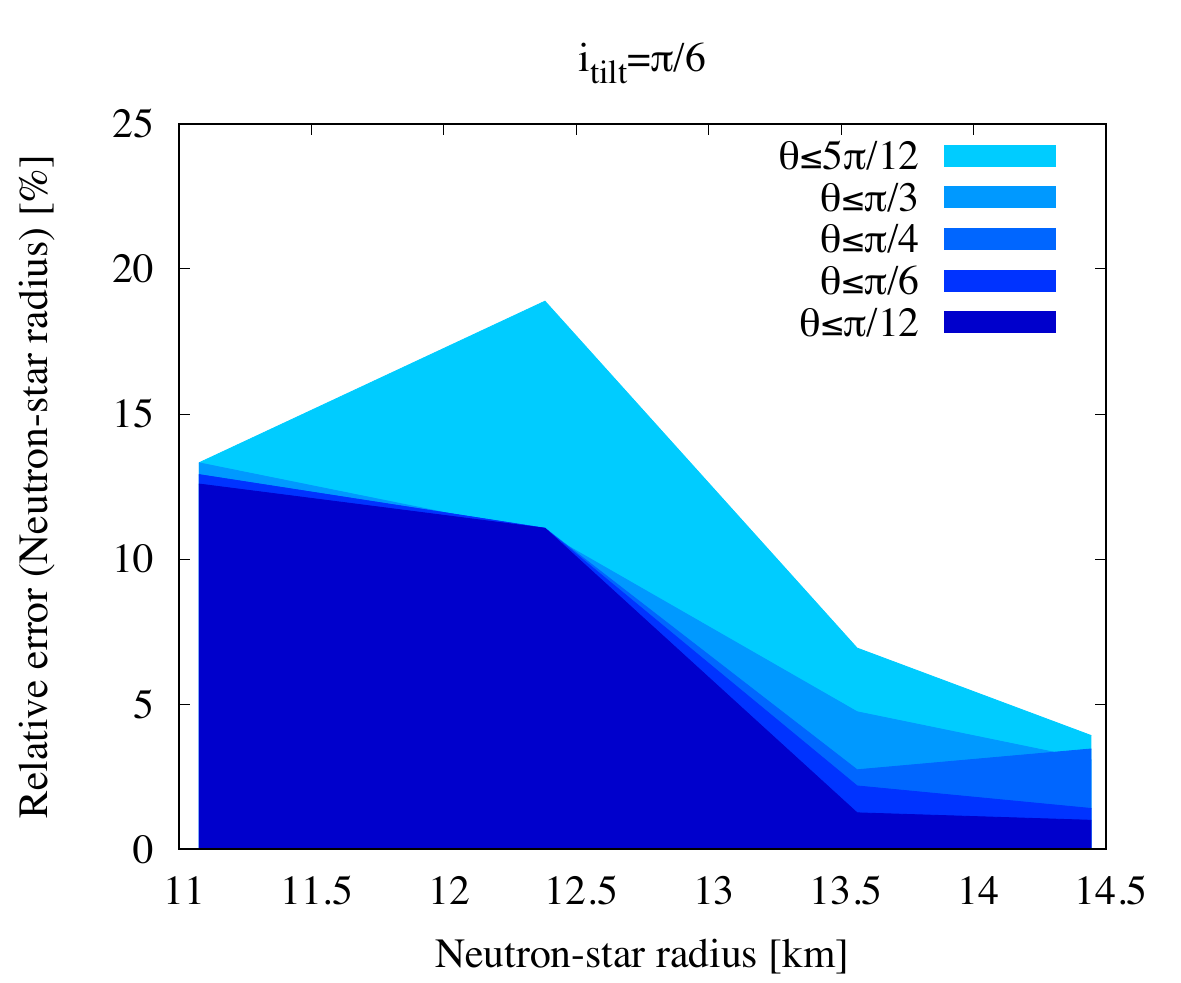}\\
	\end{center}
	\caption{Upper panels: The cutoff frequency measured from the reconstructed spectrum. The horizontal axis denotes the neutron-star radius. The left and right panels show the results for the case that $i_{\rm tilt}=0$ (non-precessing) and $\pi/6$ (precessing), respectively. The light green dashed lines denote the cutoff frequencies measured from the spectra of gravitational waves reconstructed from the strains observed from $\theta=0$. The filled areas show the ranges of cutoff frequencies measured from the reconstructed spectra for a given range of $\theta$. Lower panels: The relative error in the measured neutron-star radius induced by the systematic error in the cutoff frequency.}
	\label{fig:res}
\end{figure*}

	We plot the cutoff frequency extracted from the reconstructed spectra and the systematic variation in measurement of the neutron-star radius in Fig.~\ref{fig:res}. In the upper two panels, the cutoff frequency is plotted as a function of the neutron-star radius. For each $i_{\rm tilt}$, we plot the results of the cases for $\theta=0$ with light green dashed lines, and show the range of measured cutoff frequency for a given range of $\theta$. The width of the range of the cutoff frequency can be interpreted as the systematic error of our method due to the variation of the observer direction. We draw the bands showing the results for each range of $\theta$ by varying $\varphi$ in $[0:2\pi)$, because we find that the systematic error depends primarily on $\theta$. In particular, while $\theta$ can be measured from the observed waveforms, degeneracy with the orbital phase makes the measurement of $\varphi$ difficult for non-precessing cases.
	
	The lower two panels show the error of the neutron-star radius induced by the systematic error of the cutoff frequency. We assume the results for $\theta=0$ as the fiducial relation between the cutoff frequency and the neutron-star radius. Then, the measurement error of the neutron-star radius  for a given range of $\theta$ is calculated as follows:  First, for each given neutron-star radius, we calculate the corresponding cutoff frequency by the fiducial relation. Next, we search the extent of the neutron-star radius that reproduces the same cutoff frequency within the systematic error. Finally, we define the error of the neutron-star radius from the largest deviation in such an extent.
	
	Generally, the systematic error in the cutoff frequency is larger for a larger value of $\theta$. This is because large values of $\theta$ enhance the contribution of sub-dominant modes to the strain, and make it difficult to remove the modulation in the spectrum. For non-precessing cases, the systematic error in the cutoff frequency is within $\sim10\%$ when $\theta\le5\pi/12$. The systematic error in the neutron-star radius is within $\approx5\%$ in these cases. 
	
	 The systematic error in the cutoff frequency also tends to be larger for larger values of $i_{\rm tilt}$. Nevertheless, for the moderately precessing cases ($i_{\rm tilt}=\pi/6$), the cutoff frequency can also be measured within $\sim10\%$ error if $\theta\le5\pi/12$. We find that there is still strong correlation between the neutron-star radius and cutoff frequency if the neutron-star radius is larger than $\approx 13.5\,{\rm km}$, and the error in the neutron-star radius is smaller than $\approx5\%$. The correlation is weaker than the non-precessing ones for the case that the neutron-star radius is smaller than $\approx 13.5\,{\rm km}$. This is because the tidal disruption of the neutron star is less likely to occur for a large value of $i_{\rm tilt}$ even for large neutron-star radii \citep{2011PhRvD..83b4005F,2015PhRvD..92b4014K}.
	 
	For $i_{\rm tilt}=\pi/3$ and $\pi/2$, the systematic error in the cutoff frequency is larger than two previous cases shown in Fig.~\ref{fig:res}  due to a large value of $i_{\rm tilt}$. Moreover, the correlation between the neutron-star radius and the cutoff frequency is much weaker than the cases with $i_{\rm tilt}=0$ and $\pi/6$. For $i_{\rm tilt}=\pi/3$, the correlation between the neutron-star radius and the cutoff frequency is still found if the neutron-star radius is larger than $\approx 13.5\,{\rm km}$, but it is hidden in the systematic error of the cutoff frequency unless $\theta\le\pi/6$. For $i_{\rm tilt}=\pi/2$, the correlation is hardly found even if $\theta=0$. Therefore, it is difficult to constrain the neutron-star radius from the cutoff frequency for these cases. 

\section{Discussion}
	In this letter, we studied the systematic error in measuring the cutoff frequency of black hole-neutron star merger using a method for extracting the face-on waveforms from the observed strain. This work is done for the case that the systematic error due to the detector noise is much smaller than the systematic error. Under this assumption, we show that the systematic error can be within $\sim10\%$ even for the case that the direction of the fiducial observer, $\theta\lesssim5\pi/12$. 
	
	Employing the data obtained by Refs.~\citep{ 2015PhRvD..92d4028K,2015PhRvD..92b4014K}, it is found that systematic error in the neutron-star radius can be within $\sim5\%$ for non-precessing cases. Furthermore, the systematic error in the neutron-star radius larger than $13.5\,{\rm km}$ can be smaller than $\sim5\%$ for moderately precessing cases. However, the extremely high signal-to-noise ratio events will not be expected in the near future, and thus, the measurability in the presence of the detector noise should be studied. We also need to extend the study to the other cases that the tidal disruption of the neutron star occurs significantly, such as  the cases with the small mass ratio or large dimensionless spin parameter. In particular, a black hole-neutron star binary with a high black-hole spin would be interesting to check if our method is applicable to the system which is largely precessing but the tidal disruption is still significant.
	
	The systematic error in the cutoff frequency might be improved by employing accurate numerical waveforms. A residual eccentricity with $\approx0.01$ is present in the initial condition of our simulation. A residual eccentricity induces an oscillatory modulation in the phase evolution, and thus, enhances the contribution of sub-dominant modes in the spectrum. Furthermore, the waveforms contain only last $10$--$17$ cycles before the merger. This shortness of the data also induces the error in the dominant modes via broadening of the peaks in the mode spectrum. Therefore, the systematic error in the cutoff frequency might be smaller than the current result if we employ data with more wave cycles and with less eccentricity. Indeed, we find that the cutoff frequency measured from various directions agrees with each other within $\approx 5\%$ if we employ binary black hole waveforms in Ref.~\citep{SXS:catalog}, which contains more wave cycles and less residual eccentricity than the ones employed in this work. We are now also working on a high-accuracy longterm simulation for a black hole-neutron star merger employing a low residual eccentricity initial data, and hence, the systematic error in the cutoff frequency would be improved from what we found in this letter.
	
\begin{acknowledgements}
This work was supported by Grant-in-Aid for Scientific Research (Grant Nos. 24244028, 16K05347, 16H02183, 16H06342, 17H01131, 14J02950) of Japanese JSPS. Kyohei Kawaguchi was supported by JSPS Postdoctoral Fellowships for Research Abroad.
\end{acknowledgements}
\bibliographystyle{unsrt}
\bibliography{ref}

\begin{thebibliography}{10}

\bibitem{2017PhRvL.118v1101A}
B.~P. {Abbott}, R.~{Abbott}, T.~D. {Abbott}, F.~{Acernese}, K.~{Ackley},
  C.~{Adams}, T.~{Adams}, P.~{Addesso}, R.~X. {Adhikari}, V.~B. {Adya}, and
  et~al.
\newblock {GW170104: Observation of a 50-Solar-Mass Binary Black Hole
  Coalescence at Redshift 0.2}.
\newblock {\em Physical Review Letters}, 118(22):221101, June 2017.

\bibitem{2015CQGra..32b4001A}
F.~{Acernese}, M.~{Agathos}, K.~{Agatsuma}, D.~{Aisa}, N.~{Allemandou},
  A.~{Allocca}, J.~{Amarni}, P.~{Astone}, G.~{Balestri}, G.~{Ballardin}, and
  et~al.
\newblock {Advanced Virgo: a second-generation interferometric gravitational
  wave detector}.
\newblock {\em Classical and Quantum Gravity}, 32(2):024001, January 2015.

\bibitem{2012CQGra..29l4007S}
K.~{Somiya}.
\newblock {Detector configuration of KAGRA-the Japanese cryogenic
  gravitational-wave detector}.
\newblock {\em Classical and Quantum Gravity}, 29(12):124007, June 2012.

\bibitem{2010CQGra..27q3001A}
J.~{Abadie}, B.~P. {Abbott}, R.~{Abbott}, M.~{Abernathy}, T.~{Accadia},
  F.~{Acernese}, C.~{Adams}, R.~{Adhikari}, P.~{Ajith}, B.~{Allen}, and et~al.
\newblock {TOPICAL REVIEW: Predictions for the rates of compact binary
  coalescences observable by ground-based gravitational-wave detectors}.
\newblock {\em Classical and Quantum Gravity}, 27(17):173001, September 2010.

\bibitem{2015ApJ...806..263D}
M.~{Dominik}, E.~{Berti}, R.~{O'Shaughnessy}, I.~{Mandel}, K.~{Belczynski},
  C.~{Fryer}, D.~E. {Holz}, T.~{Bulik}, and F.~{Pannarale}.
\newblock {Double Compact Objects III: Gravitational-wave Detection Rates}.
\newblock {\em \apj}, 806:263, June 2015.

\bibitem{2015MNRAS.448..928K}
C.~{Kim}, B.~B.~P. {Perera}, and M.~A. {McLaughlin}.
\newblock {Implications of PSR J0737-3039B for the Galactic NS-NS binary merger
  rate}.
\newblock {\em mnras}, 448:928--938, March 2015.

\bibitem{2016ARNPS..66...23F}
R.~{Fern{\'a}ndez} and B.~D. {Metzger}.
\newblock {Electromagnetic Signatures of Neutron Star Mergers in the Advanced
  LIGO Era}.
\newblock {\em Annual Review of Nuclear and Particle Science}, 66:23--45,
  October 2016.

\bibitem{2006PhRvD..74l1503S}
M.~{Shibata} and K.~{Ury{\=u}}.
\newblock {Merger of black hole-neutron star binaries: Nonspinning black hole
  case}.
\newblock {\em \prd}, 74(12):121503, December 2006.

\bibitem{2007CQGra..24S.125S}
M.~{Shibata} and K.~{Uryu}.
\newblock {Merger of black hole neutron star binaries in full general
  relativity}.
\newblock {\em Classical and Quantum Gravity}, 24:S125--S137, June 2007.

\bibitem{2008PhRvD..77h4015S}
M.~{Shibata} and K.~{Taniguchi}.
\newblock {Merger of black hole and neutron star in general relativity: Tidal
  disruption, torus mass, and gravitational waves}.
\newblock {\em \prd}, 77(8):084015, April 2008.

\bibitem{2008PhRvD..77h4002E}
Z.~B. {Etienne}, J.~A. {Faber}, Y.~T. {Liu}, S.~L. {Shapiro}, K.~{Taniguchi},
  and T.~W. {Baumgarte}.
\newblock {Fully general relativistic simulations of black hole-neutron star
  mergers}.
\newblock {\em \prd}, 77(8):084002, April 2008.

\bibitem{2008PhRvD..78j4015D}
M.~D. {Duez}, F.~{Foucart}, L.~E. {Kidder}, H.~P. {Pfeiffer}, M.~A. {Scheel},
  and S.~A. {Teukolsky}.
\newblock {Evolving black hole-neutron star binaries in general relativity
  using pseudospectral and finite difference methods}.
\newblock {\em \prd}, 78(10):104015, November 2008.

\bibitem{2012PhRvD..85l7502S}
M.~{Shibata}, K.~{Kyutoku}, T.~{Yamamoto}, and K.~{Taniguchi}.
\newblock {Erratum and Addendum: Gravitational waves from black hole-neutron
  star binaries: Classification of waveforms}.
\newblock {\em \prd}, 85(12):127502, June 2012.

\bibitem{2009PhRvD..79d4024E}
Z.~B. {Etienne}, Y.~T. {Liu}, S.~L. {Shapiro}, and T.~W. {Baumgarte}.
\newblock {General relativistic simulations of black-hole-neutron-star mergers:
  Effects of black-hole spin}.
\newblock {\em \prd}, 79(4):044024, February 2009.

\bibitem{2010CQGra..27k4106D}
M.~D. {Duez}, F.~{Foucart}, L.~E. {Kidder}, C.~D. {Ott}, and S.~A. {Teukolsky}.
\newblock {Equation of state effects in black hole-neutron star mergers}.
\newblock {\em Classical and Quantum Gravity}, 27(11):114106, June 2010.

\bibitem{2010PhRvD..82d4049K}
K.~{Kyutoku}, M.~{Shibata}, and K.~{Taniguchi}.
\newblock {Gravitational waves from nonspinning black hole-neutron star
  binaries: Dependence on equations of state}.
\newblock {\em \prd}, 82(4):044049, August 2010.

\bibitem{2011PhRvD..84f4018K}
K.~{Kyutoku}, H.~{Okawa}, M.~{Shibata}, and K.~{Taniguchi}.
\newblock {Gravitational waves from spinning black hole-neutron star binaries:
  dependence on black hole spins and on neutron star equations of state}.
\newblock {\em \prd}, 84(6):064018, September 2011.

\bibitem{2013CQGra..30m5004L}
G.~{Lovelace}, M.~D. {Duez}, F.~{Foucart}, L.~E. {Kidder}, H.~P. {Pfeiffer},
  M.~A. {Scheel}, and B.~{Szil{\'a}gyi}.
\newblock {Massive disc formation in the tidal disruption of a neutron star by
  a nearly extremal black hole}.
\newblock {\em Classical and Quantum Gravity}, 30(13):135004, July 2013.

\bibitem{2015PhRvD..92d4028K}
K.~{Kyutoku}, K.~{Ioka}, H.~{Okawa}, M.~{Shibata}, and K.~{Taniguchi}.
\newblock {Dynamical mass ejection from black hole-neutron star binaries}.
\newblock {\em \prd}, 92(4):044028, August 2015.

\bibitem{2010PhRvL.105k1101C}
S.~{Chawla}, M.~{Anderson}, M.~{Besselman}, L.~{Lehner}, S.~L. {Liebling},
  P.~M. {Motl}, and D.~{Neilsen}.
\newblock {Mergers of Magnetized Neutron Stars with Spinning Black Holes:
  Disruption, Accretion, and Fallback}.
\newblock {\em Physical Review Letters}, 105(11):111101, September 2010.

\bibitem{2012PhRvD..85f4029E}
Z.~B. {Etienne}, Y.~T. {Liu}, V.~{Paschalidis}, and S.~L. {Shapiro}.
\newblock {General relativistic simulations of black-hole-neutron-star mergers:
  Effects of magnetic fields}.
\newblock {\em \prd}, 85(6):064029, March 2012.

\bibitem{2012PhRvD..86h4026E}
Z.~B. {Etienne}, V.~{Paschalidis}, and S.~L. {Shapiro}.
\newblock {General-relativistic simulations of black-hole-neutron-star mergers:
  Effects of tilted magnetic fields}.
\newblock {\em \prd}, 86(8):084026, October 2012.

\bibitem{2013ApJ...776...47D}
M.~B. {Deaton}, M.~D. {Duez}, F.~{Foucart}, E.~{O'Connor}, C.~D. {Ott}, L.~E.
  {Kidder}, C.~D. {Muhlberger}, M.~A. {Scheel}, and B.~{Szilagyi}.
\newblock {Black Hole-Neutron Star Mergers with a Hot Nuclear Equation of
  State: Outflow and Neutrino-cooled Disk for a Low-mass, High-spin Case}.
\newblock {\em \apj}, 776:47, October 2013.

\bibitem{2014PhRvD..90b4026F}
F.~{Foucart}, M.~B. {Deaton}, M.~D. {Duez}, E.~{O'Connor}, C.~D. {Ott},
  R.~{Haas}, L.~E. {Kidder}, H.~P. {Pfeiffer}, M.~A. {Scheel}, and
  B.~{Szilagyi}.
\newblock {Neutron star-black hole mergers with a nuclear equation of state and
  neutrino cooling: Dependence in the binary parameters}.
\newblock {\em \prd}, 90(2):024026, July 2014.

\bibitem{2011PhRvD..83b4005F}
F.~{Foucart}, M.~D. {Duez}, L.~E. {Kidder}, and S.~A. {Teukolsky}.
\newblock {Black hole-neutron star mergers: Effects of the orientation of the
  black hole spin}.
\newblock {\em \prd}, 83(2):024005, January 2011.

\bibitem{2013PhRvD..87h4006F}
F.~{Foucart}, M.~B. {Deaton}, M.~D. {Duez}, L.~E. {Kidder}, I.~{MacDonald},
  C.~D. {Ott}, H.~P. {Pfeiffer}, M.~A. {Scheel}, B.~{Szilagyi}, and S.~A.
  {Teukolsky}.
\newblock {Black-hole-neutron-star mergers at realistic mass ratios: Equation
  of state and spin orientation effects}.
\newblock {\em \prd}, 87(8):084006, April 2013.

\bibitem{2015PhRvD..92f4034K}
K.~{Kiuchi}, Y.~{Sekiguchi}, K.~{Kyutoku}, M.~{Shibata}, K.~{Taniguchi}, and
  T.~{Wada}.
\newblock {High resolution magnetohydrodynamic simulation of black hole-neutron
  star merger: Mass ejection and short gamma ray bursts}.
\newblock {\em \prd}, 92(6):064034, September 2015.

\bibitem{2015PhRvD..92b4014K}
K.~{Kawaguchi}, K.~{Kyutoku}, H.~{Nakano}, H.~{Okawa}, M.~{Shibata}, and
  K.~{Taniguchi}.
\newblock {Black hole-neutron star binary merger: Dependence on black hole spin
  orientation and equation of state}.
\newblock {\em \prd}, 92(2):024014, July 2015.

\bibitem{2015ApJ...806L..14P}
V.~{Paschalidis}, M.~{Ruiz}, and S.~L. {Shapiro}.
\newblock {Relativistic Simulations of Black Hole-Neutron Star Coalescence: The
  Jet Emerges}.
\newblock {\em \apjl}, 806:L14, June 2015.

\bibitem{2009PhRvD..79d4030S}
M.~{Shibata}, K.~{Kyutoku}, T.~{Yamamoto}, and K.~{Taniguchi}.
\newblock {Gravitational waves from black hole-neutron star binaries:
  Classification of waveforms}.
\newblock {\em \prd}, 79(4):044030, February 2009.

\bibitem{2015PhRvD..92h1504P}
F.~{Pannarale}, E.~{Berti}, K.~{Kyutoku}, B.~D. {Lackey}, and M.~{Shibata}.
\newblock {Gravitational-wave cutoff frequencies of tidally disruptive neutron
  star-black hole binary mergers}.
\newblock {\em \prd}, 92(8):081504, October 2015.

\bibitem{2008PhRvD..77b1502F}
{\'E}.~{\'E}. {Flanagan} and T.~{Hinderer}.
\newblock {Constraining neutron-star tidal Love numbers with gravitational-wave
  detectors}.
\newblock {\em \prd}, 77(2):021502, January 2008.

\bibitem{2011PhRvD..83h4051V}
J.~{Vines}, {\'E}.~{\'E}. {Flanagan}, and T.~{Hinderer}.
\newblock {Post-1-Newtonian tidal effects in the gravitational waveform from
  binary inspirals}.
\newblock {\em \prd}, 83(8):084051, April 2011.

\bibitem{2010PhRvD..81h4016D}
T.~{Damour} and A.~{Nagar}.
\newblock {Effective one body description of tidal effects in inspiralling
  compact binaries}.
\newblock {\em \prd}, 81(8):084016, April 2010.

\bibitem{2012PhRvD..85l4034B}
D.~{Bini}, T.~{Damour}, and G.~{Faye}.
\newblock {Effective action approach to higher-order relativistic tidal
  interactions in binary systems and their effective one body description}.
\newblock {\em \prd}, 85(12):124034, June 2012.

\bibitem{2014PhRvD..90l4037B}
D.~{Bini} and T.~{Damour}.
\newblock {Gravitational self-force corrections to two-body tidal interactions
  and the effective one-body formalism}.
\newblock {\em \prd}, 90(12):124037, December 2014.

\bibitem{2015PhRvL.114p1103B}
S.~{Bernuzzi}, A.~{Nagar}, T.~{Dietrich}, and T.~{Damour}.
\newblock {Modeling the Dynamics of Tidally Interacting Binary Neutron Stars up
  to the Merger}.
\newblock {\em Physical Review Letters}, 114(16):161103, April 2015.

\bibitem{2012PhRvD..85d4061L}
B.~D. {Lackey}, K.~{Kyutoku}, M.~{Shibata}, P.~R. {Brady}, and J.~L.
  {Friedman}.
\newblock {Extracting equation of state parameters from black hole-neutron star
  mergers: Nonspinning black holes}.
\newblock {\em \prd}, 85(4):044061, February 2012.

\bibitem{2014PhRvD..89d3009L}
B.~D. {Lackey}, K.~{Kyutoku}, M.~{Shibata}, P.~R. {Brady}, and J.~L.
  {Friedman}.
\newblock {Extracting equation of state parameters from black hole-neutron star
  mergers: Aligned-spin black holes and a preliminary waveform model}.
\newblock {\em \prd}, 89(4):043009, February 2014.

\bibitem{1994PhRvD..49.6274A}
T.~A. {Apostolatos}, C.~{Cutler}, G.~J. {Sussman}, and K.~S. {Thorne}.
\newblock {Spin-induced orbital precession and its modulation of the
  gravitational waveforms from merging binaries}.
\newblock {\em \prd}, 49:6274--6297, June 1994.

\bibitem{1995PhRvD..52..821K}
L.~E. {Kidder}.
\newblock {Coalescing binary systems of compact objects to
  (post)$^{5/2}$-Newtonian order. V. Spin effects}.
\newblock {\em \prd}, 52:821--847, July 1995.

\bibitem{2011PhRvD..84b4046S}
P.~{Schmidt}, M.~{Hannam}, S.~{Husa}, and P.~{Ajith}.
\newblock {Tracking the precession of compact binaries from their
  gravitational-wave signal}.
\newblock {\em \prd}, 84(2):024046, July 2011.

\bibitem{2012PhRvD..86j4063S}
P.~{Schmidt}, M.~{Hannam}, and S.~{Husa}.
\newblock {Towards models of gravitational waveforms from generic binaries: A
  simple approximate mapping between precessing and nonprecessing inspiral
  signals}.
\newblock {\em \prd}, 86(10):104063, November 2012.

\bibitem{2017arXiv170507459K}
K.~{Kawaguchi}, K.~{Kyutoku}, H.~{Nakano}, and M.~{Shibata}.
\newblock {Extracting the orbital axis from gravitational waves of precessing
  binary systems}.
\newblock {\em ArXiv e-prints}, May 2017.

\bibitem{2011CQGra..28l5023S}
B.~F. {Schutz}.
\newblock {Networks of gravitational wave detectors and three figures of
  merit}.
\newblock {\em Classical and Quantum Gravity}, 28(12):125023, June 2011.

\bibitem{SXS:catalog}
\url{http://www.black-holes.org/waveforms}.

\end{thebibliography}
\end{document}